\documentstyle[eqsecnum,aps]{revtex}
\draft
\begin{document}

\preprint{IC/96/115 \\ BA--96--28}

\title{ Supersymmetric Inflation \\
With Constraints on Superheavy Neutrino Masses }

\author{G. Lazarides}

\address{Physics Division, School of Technology, University of Thessaloniki, \\
Thessaloniki, Greece}

\author{ R.K. Schaefer and Q. Shafi }

\address{Bartol Research Institute, University of Delaware,\\ Newark, DE, 
19716}
\date{\today}
\maketitle
\begin{abstract}

We consider a supersymmetric model of inflation in which the primordial density
fluctuations are nearly scale invariant (spectral index $n \approx 0.98$) 
with amplitude proportional to $(M/M_{Planck})^2$, where $M\sim 10^{16}$ GeV
denotes the scale of the gauge symmetry breaking associated with inflation. 
The 60 or so e-foldings take place when all relevant scales are close to $M$, 
which helps suppress supergravity corrections. The gravitino and 
baryogenesis (via leptogenesis) constraints help determine the two heaviest 
right handed neutrino masses to be $\approx$ $2\times 10^{13}$ GeV and 
$6\times 10^{9}$ GeV.  
\end{abstract}


\narrowtext

The apparent existence of the supersymmetric grand unification scale of
$M_{GUT} (\sim
10^{16}$ GeV), which is hinted at by both theory and an extrapolation of
the LEP data,
suggests that the `small' ratio $M_{GUT}/M_{P} \sim 10^{-3}$ (where
$M_P=1.22 \times
10^{19}$ GeV is the Planck mass) may play an important role in particle
physics and 
cosmology. From the viewpoint of inflationary cosmology, in particular, it
seems desirable
to have a scenario in which  primordial density fluctuations could be
related to the above ratio. Moreover, if all scales associated with the
relevant inflationary phase are close to $M_{GUT}$, then we are (more or less) 
assured that the supergravity corrections are adequately suppressed.

A step in this direction was recently taken \cite{dvali94} when it was
realized that, within the framework of relatively simple supersymmetric 
models, a `hybrid' inflationary scenario \cite{linde} can be implemented with 
a number of remarkable features. In particular, the primordial density 
fluctuations are essentially scale invariant (scalar spectral index $n
\simeq 0.98$) with magnitude proportional to $(M/M_{P})^2$, where $M$
denotes the mass scale of the associated gauge symmetry breaking 
$(G \rightarrow H)$.  Cosmic background temperature anisotropy data constrain 
this scale to be $M \approx (5-6) \times 10^{15}$ GeV which strongly suggests 
embedding the model in a suitable grand unified scheme (see remarks below). 
Two other features of this scheme are worth emphasizing:

\begin{list}%
\setlength{\rightmargin=0cm}{\leftmargin=0cm}
\item[{\bf i.}] The inflationary phase is `driven' by radiative corrections
which result from
supersymmetry breaking in the very early universe;

\item[{\bf ii.}] The phase transition from $G \rightarrow H$ occurs at the
end of the inflationary
epoch so that this symmetry breaking should not produce monopoles.
In other words, this inflationary scenario will not work for the minimal
$SU(5)$ model, even if the scale $M$ had turned out to be precisely the SUSY 
GUT scale. 
\end{list}

In this letter we want to be quite specific about what $G$ and $H$ are,
although a
detailed discussion on how they are embedded in a supersymmetric grand
unified theory
such as $SO(10)$ or $SU(3)_c \times SU(3)_L \times SU(3)_R$, $((SU(3))^3$ for
short) will not be attempted. We will take $G$ to be the subgroup $SU(3)_c
\times
SU(2)_L \times SU(2)_R \times U(1)_{B-L}$ of these two groups, such that
the scale $M$
is associated with the breaking of $SU(2)_R \times U(1)_{B-L}$ to $U(1)_Y$.
We will
explore, in particular, the `reheat' phase that follows inflation. The
gravitino constraint on
the ``reheat" temperature, $T_{R}$, leads to important constraints on the
masses of the
heavy `right
handed' neutrinos. In turn, light neutrino masses $m_{\nu_{\tau}} \sim 4$
eV and
$m_{\nu_{\mu}} \sim 10^{-2.8}$ eV fit nicely into the scheme, and the
observed baryon
asymmetry of the universe can be produced via a primordial lepton asymmetry
resulting
from the decay of right handed neutrinos. We end with a brief comparison of
the resulting
cold plus hot dark matter scenario with observations.

We begin with the following globally supersymmetric renormalizable
superpotential $W$ \cite{inflsup}:
\begin{equation}
W\ =\ \kappa S \bar{\phi} \phi \ -\ \mu^2 S\ \ (\kappa>0,\ \mu>0), 
\end{equation}
where $\phi,\ \bar{\phi}$ denote the standard model singlet components of a
conjugate
pair of $SU(2)_R \times U(1)_{B-L}$ doublet left handed superfields, and
$S$ is a gauge
singlet left handed superfield. An R-symmetry, under which $S \rightarrow
e^{i \alpha} S,\
\bar{\phi} \phi \rightarrow \bar{\phi} \phi$, and $W
\rightarrow e^{i \alpha} W$, can ensure that the rest of the renormalizable
terms are either absent or irrelevant. Note that the
gauge quantum numbers of $\phi$ are precisely the same as the ones of the
`matter'
right handed neutrinos. But they are distinct (!) superfields and, in
particular, the latter do
not have the conjugate partners. From $W$, one writes down the potential
$V$ as a
function of the scalar fields $\phi, \bar{\phi}, S$:
\begin{eqnarray}
V(\phi, \bar{\phi}, S)\ &=& \ \kappa^2 \mid S \mid^2 \ [\ \mid \phi \mid^2  +
 \mid \bar{\phi}
\mid^2\ ] \nonumber \\ 
&+& \mid \kappa \phi \bar{\phi} - \mu^2 \mid^2  +  D-{\rm terms}.
\end{eqnarray}
The D-terms
vanish along the D-flat direction $\phi = \bar{\phi}^*$ which contains the
supersymmetric
minimum
\begin{equation}
\begin{array}{ccl}
\langle S\rangle & = & 0, \\
\\
\langle \mid \phi \mid \rangle & = & \langle \mid \bar{\phi} \mid \rangle =
\mu/\sqrt{\kappa}
\equiv M.
\end{array}
\end{equation}
Using an appropriate R-transformation, $S$ can be brought to the real axis,
i.e.,
$S=\sigma/\sqrt{2}$, where $\sigma$ is a normalized real scalar field.

The important point now is that in the early universe the scalar fields are
displaced from
the above minimum. In particular, for $S > S_c = M$, the potential $V$ is
minimized by
$\phi = \bar{\phi} = 0$. The energy density is dominated by $\mu^4$ which
therefore
leads to an exponentially expanding inflationary phase (hybrid inflation).
As emphasized
in \cite{dvali94}, there are important radiative corrections under these
conditions \cite{soft}. At one
loop, and for $S$ sufficiently larger than $S_c$, the inflationary
potential is given by
\begin{equation}
V_{eff} (S) = \mu^4 \left[ 1 + \frac{\kappa^2}{16\pi^2} \left( ln
\left(\frac{\kappa^2
S^2}{\Lambda^2}\right) + \frac{3}{2} - \frac{S_c^4}{12S^4} + \cdots \right)
\right].
\label{veff}
\end{equation}

Using equation (\ref{veff}), one finds \cite{liddle93} that 
the cosmic microwave quadrupole anisotropy amplitude
$(\Delta T/T)_Q \approx 8 \pi (N_Q/45)^{1/2} (x_Q/y_Q) 
 (M/M_{P})^2$, and the primordial density fluctuation spectral
index $n \simeq 0.98$. Here
$N_Q \approx 50 - 60$ denotes the relevant number of e-foldings experienced
by the
universe between the time the quadrupole scale exited the horizon and the
end of
inflation, $y_Q=
x_Q(1-7/(12x_Q^2)
+\cdots)$ with $x_Q = S_Q/M$, and $S_Q$ is the value of the scalar
field $S$ when the scale which evolved to the present horizon size crossed
outside the
de Sitter horizon during inflation.
 Also from equation (\ref{veff}), one finds $\kappa \approx \frac{8
\pi^{3/2}}{\sqrt{N_Q}} \ y_Q \left(\frac{M}{M_P} \right) $.

The inflationary phase ends as $S$ approaches $S_c$ from above. Write $S =
x S_c$,
where $x=1$ corresponds to the phase transition from $G \rightarrow H$
which,  it
turns out, more or less coincides with the end of the inflationary phase
(this is checked by
noting the amplitude of the quantities
$\epsilon = \frac{M_P^2}{16 \pi} (V^{\prime}/V)^2$ and $\eta =
\frac{M_P^2}{8 \pi} (V^{\prime
\prime}/V)$, where the prime refers to derivatives with respect to the
field $\sigma$). Indeed, the
$50-60$ e-foldings needed for the inflationary scenario can be realized
even with $x
\approx 2$. An important consequence of this is that with $S \sim 10^{16}$
GeV, the supergravity corrections are negligible \cite{sugr}.

In order to estimate the `reheat' temperature we take account of the fact
that the inflaton
consists of the two complex scalar fields $S$ and $\theta=(\delta \phi + \delta
\bar{\phi})/\sqrt{2}$, where $\delta \phi = \phi - M$, $\delta \bar{\phi} =
\bar{\phi} - M$, with
mass $m_{infl} = \sqrt{2}\kappa M$. We mainly concentrate on the decay
of $\theta$. Its relevant coupling to `matter' is provided by the
non-renormalizable
superpotential coupling (in symbolic form): \begin{equation}
\frac{1}{2}\left( \frac{M_{\nu^c}}{M^2} \right) \bar{\phi} \bar{\phi} \nu^c
\nu^c,
\end{equation}
where $M_{\nu^c}$ denotes the Majorana mass of the relevant right handed
neutrino
$\nu^c$. Without loss of generality we assume that the Majorana mass matrix
of the right
handed neutrinos has been brought to diagonal form with positive entries.
Clearly,
$\theta$ decays predominantly into the heaviest right handed neutrino
permitted by
phase space. (The field $S$ can rapidly decay into higgsinos through the
renormalizable
superpotential term $\xi S h^{(1)} h^{(2)}$ allowed by the gauge symmetry,
where $h^{(1)},\
h^{(2)}$ denote the electroweak higgs doublets which couple to the up and
down type
quarks respectively, and $\xi$ is a suitable coupling constant. Note that
after supersymmetry
breaking, $\langle S\rangle\sim M_S$, where $M_S\sim$ TeV denotes the
magnitude of the
breaking.)

Following standard procedures (we will soon comment on the issue of parametric
resonance), and assuming the MSSM spectrum, the `reheat' temperature $T_R$ is
given by
\begin{equation}
T_R\ \approx \frac{1}{7} \left( \Gamma_\theta M_P\right)^{1/2}, \label{reheat1}
\end{equation}
where $\Gamma_\theta \approx (1/ 16\pi) (\sqrt{2} M_{\nu^c}/M)^2 \sqrt{2}
\kappa M$ is the decay rate of $\theta$. Substituting $\kappa$ as a function 
of $N_Q$, $y_Q$, and $M$, we find
\begin{equation}
T_R \approx \ {1\over 12} \left( \frac{56}{N_Q} \right)^{1/4} \sqrt{y_Q}\
M_{\nu^c}.
\label{reheat2}
\end{equation}

Several comments are in order:
\begin{list}%
\setlength{\rightmargin=0cm}{\leftmargin=0cm}
\item[{\bf i.}] For $x_Q$ on the order of unity the `reheat' temperature is
essentially determined
by the mass of the heaviest right handed neutrino the inflaton can decay into;

\item[{\bf ii.}] The well known gravitino problem requires that $T_R$ lie
below $10^8-10^{10}$
GeV, unless a source of late stage entropy production is available. Given the
uncertainties, we will interpret the gravitino constraint as the
requirement that $T_R
\stackrel{_<}{_\sim} 10^{9}$ GeV.

\item[{\bf iii.}] In deriving equation (\ref{reheat2}) we have ignored the
phenomenon of parametric resonance. This is justified because the 
oscillation amplitude is of order $M$ (not $M_P$!), such that the induced
scalar mass ($\sim M_{\nu^c}$) is smaller than the inflaton mass
$\sqrt{2} \kappa M$.  Note that here $M_{\nu^c}$ denotes the mass of the
heaviest right handed neutrino super-multiplet the inflaton can decay into.  
\end{list}

To proceed further we will need some details from the see-saw mechanism for the
generation of light neutrino masses. For simplicity, we will ignore the
first family of quarks
and leptons. The Majorana mass matrix of the right handed neutrinos can then be
brought (by an appropriate unitary transformation on the right handed
neutrinos) to the
diagonal form with real positive entries \begin{equation}
{\cal M} = \left(\begin{array}{cc} M_1 & 0 \\ 0& M_2 \end{array} \right)\ \
\ (M_1,\ M_2>0).
\label{Mmatrix}
\end{equation}
An appropriate unitary rotation can then be further performed on the left
handed
neutrinos so that the
(approximate) see-saw light neutrino mass matrix $m_D {\cal M}^{-1}
\tilde{m}_D$, $m_D$
being the neutrino Dirac matrix, takes the diagonal form \begin{equation}
m_D\frac{1}{\cal M}\tilde{m}_D = \left(\begin{array}{cc} m_1 & 0 \\ 0 & m_2
\end{array}
\right).
\end{equation}
($m_1,\ m_2$ are, in general, complex) \cite{buchm90}. In this basis of
right and left
handed neutrinos, the elements of
\begin{equation}
m_D = \left(\begin{array}{cc} a & b \\ c& d \end{array} \right),
\label{diracmatrix}
\end{equation}
are not all independent. They can be expressed in terms of only three complex
parameters $a,\ d$, and $\eta$, where $\eta = - [M_1/M_2]^{1/2}(b/a) = [M_2
/M_1]^{1/2}(c/d)$. 

We will now assume that $m_D$ coincides asymptotically (at the SUSY GUT scale
$M_{GUT} \simeq 2\times 10^{16}$ GeV) with the up type quark mass matrix as
is the
case in many GUT models. 
Restricting ourselves, from now on, to the case
where $|\eta|\sim 1$ and $M_1/M_2 \gg 1$, we have $|a|\gg |b|$ and $|c| \gg
|d|$. Without
much loss of generality we can further take $|c| \ll |a|$ so that $a$ is
the dominant
element in $m_D$. In fact, one can numerically show that the primordial lepton
asymmetry of the universe (see below) is maximized in this region of the
parameter space.  Under these assumptions the asymptotic top and charm masses 
are $|m_t| \approx |a|$ and $|m_c| \approx |d|\ |1+\eta^2|$. Since 
$|m_2|\ll |m_1|$, we can make the following identification of the light 
neutrino mass eigenstates
\begin{equation}
m_{\nu_\tau} = |m_1|={|a|^2 \over M_1}|1+\eta^2|,\ \ m_{\nu_\mu} = |m_2|=
{|d|^2\over M_2}|1+\eta^2|. \label{mnus}
\end{equation}
We can then get the useful relations
\begin{equation}
M_2 \approx {m_c^2 m_t^2\over m_{\nu_\mu} m_{\nu_\tau}} {1\over M_1},\ \ \ \
|1+\eta^2|\approx {m_{\nu_\tau}\over m_t^2} M_1. \label{mscales}
\end{equation}

We are now ready to draw some important conclusions concerning neutrino
masses that
are more or less model independent. Assuming that the inflaton
predominantly decays to
the heaviest right handed neutrino ({\it i.e.} $M_{\nu^c}= M_1$ in equation
(\ref{reheat2}))
and employing condition (ii), we obtain $M_1 \stackrel{_<}{_\sim}
9.3\times 10^9$ GeV
for $N_Q\approx 56$ and $x_Q\approx 2$. Equation (\ref{mnus})
then implies 
an unacceptably large $m_{\nu_\tau}$ for $|\eta|\sim 1$. Thus, we are led
to our first
important conclusion: the inflaton should decay to the second heaviest
right handed
neutrino and consequently $M_{\nu^c} = M_2$ in equation (\ref{reheat2}).
Combining
this equation with equation (\ref{mscales}) we obtain
\begin{equation}
T_R \approx {1\over 12}\left({56\over N_Q}\right)^{1/4} {m_c^2 m_t^2\over
m_{\nu_\mu}
m_{\nu_\tau}} {y_Q^{1/2} \over M_1} \approx 9.2 \times 10^{21}\ {y_Q^{1/2}
\over M_1}\
{\rm GeV.}
\label{reheat3}
\end{equation}
Here we put $N_Q=56$ which is easily justifiable by standard methods at the
end of the
calculation after having fixed the values of all relevant parameters. Also,
we took $m_t =
110$ GeV, $m_c=0.24$ GeV, which are consistent with the assumption that below
$M_{GUT}$ the theory reduces to MSSM with large $\tan\beta$~\cite{babu}.
Moreover, we took $m_{\nu_\mu} \approx 10^{-2.8}$ eV which
lies at the center of the region consistent with the resolution of the
neutrino solar puzzle
via the small angle MSW mechanism. The value $m_{\nu_\tau} \approx 4$ eV  is
consistent with the light tau neutrino playing an essential role in the
formation of large scale structure in the universe. 

The value of $M_1$ is restricted by the fact that the inflaton should not
decay to the
corresponding right handed `tau' neutrino
\begin{eqnarray}
M_1 \geq {m_{infl}\over 2}= {\kappa M\over \sqrt{2}} &\approx& \left( {45
\pi \over
2}\right)^{1/2} {y_Q^2 \over N_Q x_Q} M_P \left( {\Delta T\over T} \right)_Q
\nonumber \\
&\approx & y_Q^2 x_Q^{-1}\ 1.2\times 10^{13}\ {\rm GeV}. \label{bigm1}
\end{eqnarray}
It is interesting to note that since the right handed neutrinos acquire
their masses from
superpotential terms $\lambda \frac{\bar{\phi}\bar{\phi} \nu^c\nu^c}{M_c}$,
where
$M_c=M_P/\sqrt{8\pi}\approx 2.4\times 10^{18}$ GeV and $\lambda
\stackrel{_<}{_\sim}
1$, $M_1=2\lambda M^2/M_c \stackrel{_<}{_\sim} (y_Q/ x_Q) 2.9 \times 10^{13}$ GeV
(for $N_Q=56$, $(\Delta T/T)_Q = 6.6\times
10^{-6}$). Thus,
from equation (\ref{bigm1}), $y_Q \stackrel{_<}{_\sim} 2.4$ which implies $x_Q
\stackrel{_<}{_\sim} 2.6$, and restricts the relevant part of inflation at
values of $S\sim 10^{16}$ GeV. 

To maximize the primordial lepton asymmetry (see below) we choose the bound in
equation (\ref{bigm1}) to be saturated. Equation (\ref{reheat3}) then gives
\begin{eqnarray}
T_R &\approx& x_Q y_Q^{-3/2}\ 7.6 \times 10^8\ \left( {\Delta T/T \over
6.6\times 10^{-6} }
\right)^{-1} \left( {N_Q \over 56 } \right)^{3/4}\nonumber \\
& &\left( {m_c \over 0.24 {\rm GeV} } {m_t \over 110 {\rm GeV} } \right)^2 
\left( {m_{\nu_\mu} \over 10^{-2.8} {\rm eV} } {m_{\nu_\tau} \over 4 {\rm eV} }
\right)^{-1} {\rm GeV},
\label{reheat4}
\end{eqnarray}
which satisfies condition (ii) for all allowed values of $y_Q$. Eq.
(\ref{mscales}) implies
\begin{equation}
\label{m2eta}
M_2 \approx x_Q y_Q^{-2}\ 9\times 10^{9}\ {\rm GeV},\ \ |1 + \eta^2|
\approx 4\ y_Q^2 x_Q^{-1}.
\label{}
\end{equation}
This implies that the errors in the asymptotic formulas for the top and charm 
masses are $<1$\%.  

The observed baryon asymmetry of the universe can be generated by first
producing a
primordial lepton asymmetry via the out-of-equilibrium decay of the right
handed
neutrinos, which emerge as decay products of the inflaton field at `reheating'
\cite{fugu86}.  It is important though to ensure that the
lepton asymmetry is not erased by lepton number violating 2-2 scatterings
at all
temperatures between $T_R$ and 100 GeV \cite{harvey90}. In our case this
requirement
is automatically satisfied since at temperatures above $10^7$ GeV the
lepton asymmetry
is protected \cite{ibanez92} by
supersymmetry, whereas at temperatures between $10^7$ and 100 GeV, as one can
easily show, these 2-2 scatterings are well out of equilibrium. The out-of
-equilibrium
condition for the decay of the right handed neutrinos is also satisfied
since $M_2 \gg
T_R$ for all relevant values of $x_Q$. The primordial lepton asymmetry is
estimated to be \cite{fugu86} 
\begin{equation}
{n_L\over s} \approx {9\over 8 \pi} {T_R\over m_{infl}} {M_2 \over M_1} {
{\rm Im}
(m_D^\dagger m_D/ |\langle h^{(1)}\rangle|^2)^2_{21} \over (m_D^\dagger
m_D/ |\langle
h^{(1)}\rangle|^2)_{22} }. \end{equation}
Equation (\ref{diracmatrix}) combined with the fact
that $|c||d| \ll |a||b|$ then gives 
\begin{equation}
{n_L\over s} \stackrel{_<}{_\sim}  {9\over 8 \pi} {T_R\over m_{infl}} 
{M_2 \over M_1} {m_t^2 \over |\langle h^{(1)}\rangle|^2}, 
\end{equation}
which, using equations (\ref{mscales}) - (\ref{m2eta}) and the fact that
$|\langle h^{(1)}\rangle|\approx 174$ GeV for large tan$\beta$, becomes
\begin{eqnarray}
{n_L \over s} &\stackrel{<}{_\sim}& x_Q^4 y_Q^{-15/2}\ 3.4 \times 10^{-9} 
\left( {\Delta T/T \over
6.6\times
10^{-6} } \right)^{-4} \left( {N_Q \over 56 } \right)^{15/4}\nonumber \\
& &\left( {m_c \over
0.24 {\rm GeV} } \right)^4 \left( {m_t \over 110 {\rm GeV} } \right)^6
\left( {m_{\nu_\mu} \over
10^{-2.8} {\rm eV} }\ {m_{\nu_\tau} \over 4 {\rm eV} }
\right)^{-2}.
\end{eqnarray}
For $x_Q\approx 2$ ($y_Q\approx 1.7$), this gives $n_L/s
\stackrel{_<}{_\sim} 10^{-9}$
which is large enough to account for the observed baryon asymmetry. Also 
$M\approx 5.47 \times  10^{15}$ GeV, 
$T_R \approx 6.8 \times 10^8$ GeV, $M_1\approx 1.75 \times 10^{13}$ GeV, 
$M_2\approx 6.2 \times 10^{9}$ GeV, and $m_{infl} \approx 3.5 
\times 10^{13}$ GeV for the same value of $x_Q$.

In supersymmetric models the lightest supersymmetric particle (LSP) is
expected to be stable and is a leading cold dark matter candidate. If we 
couple this with a tau neutrino of
mass $\sim 2-6$ eV we are led to the well tested 
cold plus hot dark matter (CHDM) model \cite{ss84} of large scale structure
formation, with a
spectral index of $n=0.98$.
This model \cite{ss84} provides a consistent picture for the formation of 
large scale structure in the universe, and was used to
correctly predict \cite{sss} the primordial cosmic background radiation
fluctuation
amplitude seen by the Cosmic Background Explorer satellite \cite{COBE}.

To summarize, among the key features of the inflationary models we have
discussed one could list the role played by radiative corrections in the early
universe, the realization of inflation at scales well below $M_P$ so that the
gravitational corrections can be adequately suppressed, and the constraints on
the two heaviest right handed neutrino masses.  The resulting cold plus hot 
dark matter combination which results is an added bonus.  
One of the remaining challenges is to embed the scheme described here within a
fully unified framework. 

\acknowledgments

One of us (Q.S.) would like to acknowledge the hospitality of ICTP during
completion of this work. The work reported here is supported in part by
Grant No.DE--FG02-91 ER 40626 from the Department of Energy (USA).


\begin{references}

\bibitem{dvali94} G. Dvali, Q. Shafi, and R. K. Schaefer, {\sl Phys. Rev.
Lett.}, {\bf 73},
(1994), 1886.

\bibitem{linde} A.D. Linde, Phys. Lett. B, {\bf 259}, 38 (1991); Phys. Rev. D, 
{\bf 49}, 748, (1994).

\bibitem{inflsup} For an earlier attempt at inflation with this superpotential,
see E.J. Copeland, A.R. Liddle, D.H. Lyth, E.D. Stewart and D. Wands, Phys.
Rev. D, {\bf49}, 6410 (1994).  Supersymmetric inflation has a long history 
and there are several
reviews available, see, {\it e.g.}, A.D. Linde, ``Particle Physics and 
Inflationary Cosmology", Harwood Academic, Switzerland (1990).  

\bibitem{soft} The soft supersymmetry breaking scalar masses can be safely
ignored.

\bibitem{liddle93} A.R. Liddle and D.H. Lyth, {\sl Phys. Rep.}, {\bf231},
1, (1993).

\bibitem{sugr} The supergravity induced inflaton mass remains exactly zero 
even if we include in $W$ all possible non-renormalizable terms involving 
$S$, $\phi$, $\overline{\phi}$ allowed by the gauge and R-symmetries, 
and with a minimal K\"ahler potential (see first reference in [3] and 
E.D. Stewart, Phys. Rev. D, {\bf51}, 6847, 1995).  The fact that $S\ll M_P$ 
plays a crucial role if higher order terms are included in the K\"ahler 
potential. The inflaton mass can then be kept $\ll H$ with only a mild tuning 
of just one parameter (G.Dvali, private communication). In many other 
inflationary scenarios one needs to adjust an infinite number of parameters! 


\bibitem{buchm90} W. Buchm\"uller and D. Wyler, {\sl Phys. Lett.}, {\bf
B249}, (1990), 458;
W. Buchm\"uller and T. Yanagida, (Baryogenesis and the Scale of B-L
Breaking) DESY-
Preprint 92-172(ISSN 0418-9833).

\bibitem{babu} K.S. Babu, private communication

\bibitem{fugu86} M. Fugugita and T. Yanagida, {\sl Phys. Lett.}, {\bf
B174}, (1986), 45; G.
Lazarides and Q. Shafi, {\sl Phys. Lett.}, {\bf B258}, (1991), 305; G.
Lazarides, C.
Panagiotakopoulos, and Q. Shafi, {\sl Phys. Lett.}, {\bf B315}, (1993),
325; L.Covi,
E.Roulet and F.Vissani, SISSA preprint 66/96/EP(IC/96/73)(hep-ph/9605319).

\bibitem{harvey90} J.A. Harvey and M.S. Turner, {\sl Phys. Rev. D}, {\bf
42} (1990), 3344.

\bibitem{ibanez92} L.E. Ib\'a\~nez and F. Quevedo, {\sl Phys. Lett.}, {\bf
B283}, (1992),
261.

\bibitem{ss84} Q. Shafi and F. W. Stecker, {\sl Phys. Rev. Lett.}, {\bf
53}, 1292, (1984). For a recent review, see A.R. Liddle, D. H. Lyth, R. K. 
Schaefer, Q. Shafi, and P.T. Viana, Mon. Not. Roy. Ast. Soc., {\bf 281}, 531,
(1996). 

\bibitem{sss} R.~K.~Schaefer, Q.~Shafi, and F.~W.~Stecker, {\sl Astrophys.
J.}, {\bf 347},
575, (1989); J.~Holtzman, {\sl Astrophys. J., Suppl.}, {\bf74}, 1, (1989).

\bibitem{COBE} C.L. Bennett, {\it et al.}, {\sl Astrophys. J. Lett.}, {\bf
464}, 1 (1996).


\end{references}
\end{document}